\documentclass{PoS}

\title{Status of \$$\backslash$psi \$ (3686), \$$\backslash$psi \$ (4040), \$$\backslash$psi \$ (4160), Y (4260), \$$\backslash$psi \$ (4415) and  X (4630) charmonia like states}

\ShortTitle{Status of charmonia like state}

\author{\speaker {Manan Shah}\\
        $^1$Department of Physics, Sardar Patel University, Vallabh Vidyanagar-388120, INDIA\\
        E-mail: \email{mnshah09@gmail.com}}

\author{Kaushal Thakkar\\
       Department of Applied Physics, S V National Institute of Technology, Surat-395007, INDIA\\
       E-mail: \email{kkt$\_$loc@ashd.svnit.ac.in}}

\author{Arpit Parmar\\
       Department of Physics, Sardar Patel University, Vallabh Vidyanagar-388120, INDIA\\
       E-mail: \email{arpitspu@yahoo.co.in}}

\author{P C Vinodkumar\\
       Department of Physics, Sardar Patel University, Vallabh Vidyanagar-388120, INDIA\\
       E-mail: \email{p.c.vinodkumar@gmail.com}}

\abstract{We examine the status of charmonia like states by looking into the behaviour of the
energy level differences and regularity in the behaviour of the leptonic decay widths of the excited charmonia states. The spectroscopic states are studied using a phenomenological Martin-like confinement potential and their radial wave functions are employed to compute the di-leptonic decay widths. Their deviations from the expected behaviour provide a clue to consider them as admixtures of the nearby S and D states. The present analysis strongly favour \$$\backslash$psi \$ (3686) as admixture of $c \bar{c}$ (2S) and $c \bar{c}$g (4.1 GeV) hybrid, \$$\backslash$psi \$ (4040) and \$$\backslash$psi \$ (4160) as admixture states of charmonia (3S, 3D) states with mixing angle \$$\backslash$theta \$ = 11$^\circ$ and 45$^\circ$ respectively. We identify Y (4260) as a pure $c \bar{c}$ (4S) state whose leptonic decay is predicted as 0.65 keV. While X(4630) is closer to the $c \bar{c}$ (6S) state. The status of \$$\backslash$psi \$ (4415) is still not clear as it does not fit to be pure or admixture state.    }

\FullConference{Xth Quark Confinement and the Hadron Spectrum,\\
        October 8-12, 2012\\
        TUM Campus Garching, Munich, Germany}

\begin{document}

\section{Introduction}
Since the discovery of \$J/$\backslash$Psi\$, large number of charmonium excited states with their masses and decay widths have been recorded experimentally \cite{PDG2010}. Though their spectra and decay properties are well studied, there exist disparities related to the decay properties of their excited states. For example, the \$$\backslash$rho-$\backslash$pi\$ puzzle related to the hadronic decays of \$$\backslash$Psi\$ (3686) compared to that of \$J/$\backslash$Psi \$ (1S) \cite{Y-Q,Kisslinger} is resolved by invoking these  higher charmonia states as admixtures of the respective $ c \bar{c}$ states with $c \bar{c}g $ hybrids \cite{Kisslinger}. Further, if we consider the experimental energy level differences and leptonic decay rates of the excited states beyond 2S of the $c \bar{c}$ $(1^{--})$ states, their deviations from the expected behaviour provide a clue to consider them as admixtures of the nearby S and D states \cite{Badalian}. In this context we examine the status of \$$\backslash$psi \$ (3686), \$$\backslash$psi \$ (4040), \$$\backslash$psi \$ (4160), Y (4260), \$$\backslash$psi \$ (4415) and  X (4630) charmonia like states by looking into the behaviour of the energy level differences of charmonia states and their experimental leptonic decay widths.

\section{Methodology}
\begin{figure}[h]
\includegraphics[width=.6\textwidth]{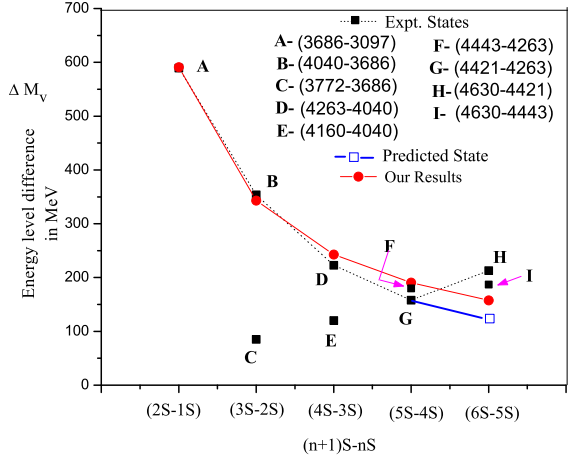}
\caption{Behavior of energy level shift of the (n+1)S$-$nS charmonium states}\label{fig:2}
\end{figure}

From the experimentally known $J^{PC}=1^{--}$ charmonium states, their energy level differences are shown in Fig (\ref{fig:2}). One of our recent theoretical predictions of these states \cite {Manan2012} are also shown for comparison. It is evident from the plot that \$$\backslash$psi \$ (3770,4160), X(4630) of charmonia like  states are off from the expected trend as seen from the graph. Looking into their leptonic decay widths similar disparities are observed for the states \$$\backslash$psi \$ (4040,4160). The disparities of the predicted higher S$-$wave masses beyond $nS (n\geq3)$ states are reported to be due to the admixture of the S$-$states with nearby D$-$states \cite{Badalian}. Thus we adopt a methodology by considering these states as disturbed with suitable mixing with the nearby states having similar parity.\\

Accordingly, the mixed state $R_{nS'} $ is represented in terms of the mixing angle $\theta$ as \cite{Badalian}
\begin{equation}\label{phi}
R_{nS'} = \cos \$ \backslash theta \$ \ R_{nS} - \sin \$\backslash theta \$ \ R_{n'D}
\end{equation}
 where the wave function at zero of the D-wave, $R_{n'D}(0)$ is defined in terms of the second derivative of the D$-$wave as $R''_{n'D}(0)/M^2_{n'D}$ \cite{Badalian}. For the admixture of $c\bar{c}g$ hybrid case, $R_{n^{'}D}$ of eqn. (\ref{phi}) is replaced by  $R_{c\bar c g}$. These disturbed wave functions at the origin are then employed to compute the leptonic decay widths of the mixed states. We have employed the predicted masses and wave functions based on a phenomenological confinement model with Martin-like potential
 \cite{Manan2012} for the present study.\\

 When we consider 50 \% admixture of $c\bar{c}g$ hybrid state bearing its mass equal to 4.1 GeV given by \cite{Iddir} yield the leptonic decay widths of \$$\backslash$psi \$ (3686) as 2.376 keV as against the predicted 1.686 keV \cite{Manan2012} which is in good agreement with the reported experimental values of 2.35 $\pm$ 0.04 keV. The mixing configuration, the mixing angle and predicted leptonic decay widths of charmonia like states of present interest are listed in Table \ref{tab100}.
%%%%%%%%%%%%%%%%%%%%%%%%%%%%%%%%%%%%%%%%%%%%%%%%%%%%%%%%%%%%%%%%%%%%%%%%%%%%%%%
\begin{figure}[h]
\includegraphics[width=.6\textwidth]{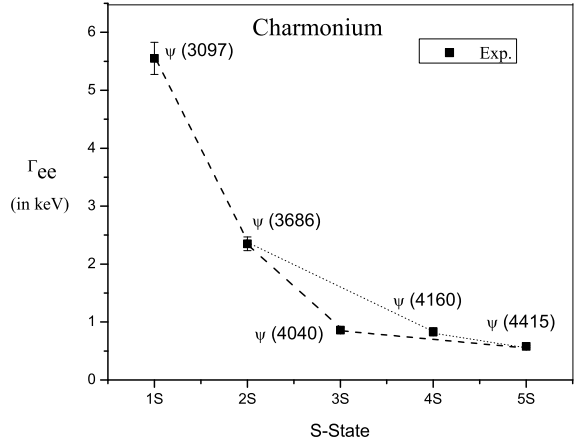}
\caption{Behavior of leptonic decay width of Exp. charmonia states}\label{fig:4}
\end{figure}
%%%%%%%%%%%%%%%%%%%%%%%%%%%%%%%%%%%%%%%%%%%%%%%%%%%%%%%%%%%%%%%%%%%%%%%%%%%%%
\begin{table}
\begin{center}
\caption{Mixing configuration and leptonic widths
(in keV) of $c\bar{c}$ states} \label{tab100}
\begin{tabular}{c|cccccccccccccc}
\hline\hline
Exp. State &  Mixed     &  \$$\backslash$theta \$  &  \$$\backslash$Gamma$^{e^+e^-}$ \$
&  \$$\backslash$Gamma$^{e^+e^-}_{[Expt.]}$ \$  & \\
& config.     &    &     &
            \cite{PDG2010} \\
\hline\hline
$\psi(3686)$ & Pure $2{^3S_1}$       &               & 1.686  & $2.35^{0.04}_{0.04}$ \\
            & ($\psi(2S),c\bar{c}g$) &   45$^\circ$  & 2.376  &        \\
\hline

$\psi(4040)$ & ($3{^3S_1}$,$3{^3D_1}$)  & 11.07$^\circ$ &  \textbf{0.896} & $0.86^{+0.07}_{-0.07}$\\
             & ($3{^3S_1}$,$2{^3D_1}$)  & 37.53$^\circ$ & 0.528   &\\
\hline
$\psi (4160)$ & $4{^3S_1}$,$2{^3D_1}$  & 46.31$^\circ$ & 0.268  & $0.48^{+0.22}_{-0.22}$ \cite{BES08}\\
              & $3{^3S_1}$,$3{^3D_1}$  & 44.62$^\circ$ & 0.398 &  \\

\hline
$Y (4260)$    & $4{^3S_1}$,$2{^3D_1}$  & 14.44$^\circ$  & 0.588  & -\\
              & $3{^3S_1}$,$3{^3D_1}$  & 69.19$^\circ$  & 0.074  & -\\
              & $4{^3S_1}$,$3{^3D_1}$  & NP             & -      & -\\

\hline
$\psi (4415)$ & $5{^3S_1}$,$3{^3D_1}$ & 32.38$^\circ$  & 0.320  & $0.58^{+0.07}_{-0.07}$\\
              & $4{^3S_1}$,$4{^3D_1}$ & 55.87$^\circ$  & 0.158  & \\
              & $5{^3S_1}$,$4{^3D_1}$ & NP             & -      &  \\
\hline
$X (4630)$    & $5{^3S_1}$,$5{^3D_1}$ & 80.23$^\circ$  & 0.005  & - \\
              & $6{^3S_1}$,$5{^3D_1}$ & 53.52$^\circ$  & 0.112  & - \\
\hline\hline

\end{tabular}

\end{center}
NP= Not Possible
\end{table}

%%%%%%%%%%%%%%%%%%%%%%%%%%%%%%%%%%%%%%%%%%%%%%%%%%%%%%%%%%%%%%%%%%%%%%%%%%%%%%%%%%%%%%%%%%%%%%%
%\begin{figure}[h]
%\includegraphics[height=2.0in,width=2.5in]{bottom.eps}
%\caption{Behavior of energy level shift of the (n+1)S$-$nS bottonium states}\label{fig:1}
%\end{figure}

\section{Results and discussion}
  Our analysis here has also provided a strong support to treat \$$\backslash$psi \$ (3686) as hybrid admixture states \cite{Kisslinger}. We find the admixture of hybrid state excludes the radiative correction to the leptonic decay widths. Further we find that \$$\backslash$psi \$ (4040) is admixture of $3^3 S_1$ and $3^3 D_1$ with mixing angle \$$\backslash$theta \$ = 11.07$^\circ$ correspond to $96.31 \%$ of $3S$ state and 3.69 $\%$ of 3D state with its leptonic decay width 0.896 keV which is in close agreement with the experimental value of $0.86 \pm 0.07 $ keV. The leptonic decay widths of \$$\backslash$psi \$ (4160) obtained here with the mixing configuration of ($3^3S_1$, $3^3D_1$) and ($4^3S_1$, $2^3D_1$) are in agreement with the experimental value $0.48 \pm 0.22$ and lie within the error bar reported by Belle Collaboration and BES \cite{Choi} but completely in disagreement with the value of $0.83 \pm 0.07$ reported by \cite{PDG2010}. Though Y (4260) can be interpreted as 4S$-$2D admixture state with mixing angle \$$\backslash$theta \$ = 14.44$^\circ$ that predicts its leptonic decay width 0.588 keV, the mixing may not be possible as the 4S and 2D masses differ by more than 200 MeV. So, we consider Y(4260) close to the $c\bar{c}$ (4S) state with predicted leptonic decay width of 0.65 keV. However  experimental determination of this width is awaited. While the state \$$\backslash$psi \$ $(4421 \pm 4)$ does not qualify to be the pure 5S state or S$-$D admixture. Experimental measurements of the leptonic decay widths of Y (4260) and X (4630) will justify their status discussed in this paper.

\section{Acknowledgments}
%\acknowledgments
The work is part of a Major research project NO. F. 40-457/2011(SR) funded by UGC.
%\end{acknowledgments}

%\ \\
%\noindent

\end{document}